\documentclass[journal,twoside,web]{ieeecolor}
\usepackage{lcsys}
\usepackage{amsmath,amsfonts}
\usepackage{algorithm}
\usepackage{algpseudocode}
\usepackage{array}
\usepackage[caption=false,font=normalsize,labelfont=sf,textfont=sf]{subfig}
\usepackage{textcomp}
\usepackage{stfloats}
\usepackage{url}
\usepackage{verbatim}
\usepackage{graphicx}
\usepackage{cite}
\usepackage{amsfonts}       
\usepackage{nicefrac}       
\usepackage{microtype}      
\usepackage{xcolor}         
\usepackage{graphicx}
\usepackage{caption}
\usepackage{booktabs}
\usepackage{amsmath}
\usepackage{amsthm}
\usepackage{cite}
\usepackage[colorlinks,bookmarks=false]{hyperref}
\hypersetup{citecolor=[RGB]{0,0,255}, urlcolor=[RGB]{0,0,255}}

\usepackage[font=footnotesize]{caption}
\usepackage{mwe}
\usepackage{mathptmx}
\usepackage{tikz}
\newcommand*\circled[1]{\tikz[baseline=(char.base)]{\node[shape=circle,draw,inner sep=0.05pt] (char) {#1};}}
\newtheorem{proposition}{Proposition}
\newtheorem{definition}{Definition}
\newtheorem{lemma}{Lemma}
\newtheorem{theorem}{Theorem}

\newtheorem{assumption}{Assumption}
\newtheorem*{design*}{Controller Design}

\renewcommand{\baselinestretch}{.98}
\allowdisplaybreaks

\DeclareMathOperator*{\argmin}{arg\,min}
\usepackage{xcolor}

\title{\LARGE \bf
Bridging Transient and Steady-State Performance in Voltage Control: A Reinforcement Learning Approach with Safe Gradient Flow}

\author{Jie Feng$^{1}$ \quad Wenqi Cui$^{2}$ \quad Jorge Cort\'es$^{3}$ \quad Yuanyuan Shi$^{1}$ 
\thanks{*The authors are supported by NSF ECCS-2200692, ECCS-2153937, and ECCS-1947050 awards and the Jacobs School Early Career Faculty Development Award.}
\thanks{$^{1}$Jie Feng and Yuanyuan Shi are with the Department of Electrical and Computer Engineering, University of California San Diego, \tt jif005@ucsd.edu, yyshi@eng.ucsd.edu.}
\thanks{$^{2}$Wenqi Cui is with the Department of Electrical and Computer Engineering, University of Washington, Seattle, \tt wenqicui@uw.edu.}
\thanks{$^{3}$Jorge Cort\'es is with the Department of Mechanical and Aerospace Engineering, University of California San Diego, \tt cortes@ucsd.edu.}%
}

\begin{document}

\maketitle
\thispagestyle{empty}
\pagestyle{empty}

\begin{abstract}
Deep reinforcement learning approaches are becoming appealing for the design of nonlinear controllers for voltage control problems, but the lack of stability guarantees hinders their deployment in real-world scenarios. This paper constructs a decentralized RL-based controller for inverter-based real-time voltage control in distribution systems. It features two components: a transient control policy and a steady-state performance optimizer. The transient policy is parameterized as a neural network, 
and the steady-state optimizer represents the gradient of the long-term operating cost function. The two parts are synthesized through a safe gradient flow framework, which prevents the violation of reactive power capacity constraints.
We prove that if the output of the transient controller is bounded and monotonically decreasing with respect to its input, then the closed-loop system is asymptotically stable and converges to the optimal steady-state solution. We demonstrate the effectiveness of our method by conducting experiments with IEEE 13-bus and 123-bus distribution system test feeders.
\end{abstract}

\section{INTRODUCTION}
Voltage safety is one of the primary concerns of power system operation, which requires the voltage magnitude to stay in an acceptable range under all working conditions~\cite{Baran1989a}. In recent years, the integration of distributed energy resources (DERs) such as roof-top solar and electric vehicles has led to rapid and unpredictable fluctuations in the load and generation profiles of the distribution systems, thus leading to challenges in real-time voltage control for distribution grids. 

There have been tremendous efforts made to overcome this challenge. Much of the attention has focused on optimizing the steady-state cost for voltage control~\cite{7436388,7028508,zhu2016fast,qu2019optimal,ZY-GC-MKS-JC:22-tps},
which refers to the operation cost 
after the system voltage has settled to equilibrium after a disturbance. However, as the system is subject to more frequent disturbances from load and generation fluctuations, optimizing the transient performance (i.e., how to optimally stabilize voltage after disturbances) becomes of equal importance.

The transient performance for the voltage control problem involves minimizing the voltage recovery time, at the minimum control effort. However, optimizing the transient cost for voltage control is a challenging task~\cite{https://doi.org/10.48550/arxiv.2209.07669}, as both the cost functions and system dynamics can be nonlinear. This is made even more challenging due to the lack of exact model knowledge and limited communications in the distribution grid.
Recently, reinforcement learning (RL) has emerged as a powerful approach for addressing model-free nonlinear control problems. There has been considerable interest in developing RL-based controllers for optimizing the transient performance of voltage control problems. We refer readers to a recent survey~\cite{9721402}.

Recent research has revealed that RL with a monotone policy network can ensure transient stability for voltage control~\cite{shi2022stability, feng2022stability, cui2022decentralized}. However, these works do not offer guarantees regarding the optimality of the steady state. Steady-state requirements are difficult to enforce in RL since training can only occur over a finite horizon. 
Motivated by the challenges, the question we want to address in this paper is,
\begin{center}
    \emph{Can RL be structured to optimize both transient and steady-state performance for voltage control?} 
\end{center}

The key idea underlying our approach is the synthesis of a 
neural-network-based transient control policy and a steady-state optimizer (represented by the gradient of the cost function) in a safe gradient flow framework~\cite{allibhoy2022control}. This enables us to 
coordinate these two sub-controllers to optimize both transient and steady-state performance while respecting the reactive power constraint and guaranteeing closed-loop stability. 
We summarize our main contributions as follows:
\begin{itemize}
    \item We design a decentralized RL-based controller that optimizes both transient and steady-state performance for the distribution system voltage control; 
    \item 
    We prove that the proposed controller design guarantees both transient stability and steady-state optimality, for strictly convex objective functions (in Theorem~\ref{thm:optimality(stability)});  
    \item We demonstrate the effectiveness of the proposed method with extensive numerical experiments. Our method can reduce over $30\%$ transient cost compared to controllers that only optimize the steady-state or transient performance, and guarantee optimal steady-state cost.
\end{itemize}


The remaining parts of this paper are organized as follows. Section \uppercase\expandafter{\romannumeral2} presents the distribution system voltage control problem and steady-state optimization solution as preliminaries. Section \uppercase\expandafter{\romannumeral3} presents the proposed transient and steady-state reinforcement learning algorithm. Its stability and optimality guarantees are rigorously proved in Section \uppercase\expandafter{\romannumeral4}. Section \uppercase\expandafter{\romannumeral5} evaluates the results in IEEE test feeders and provides a discussion, followed by concluding remarks  in Section~\uppercase\expandafter{\romannumeral6}.

\section{Model \&  Preliminaries}
In this section, we review the distribution system power flow model and introduce the voltage control problem. 
\subsection{Branch Flow Model}
We consider the linearized branch flow model~\cite{7028508} in a tree-structured distribution network for theoretical analysis. The system is defined as $\mathcal{G} = (\mathcal{N}_0,\mathcal{E})$, consisting of a set of  nodes $\mathcal{N}_0 = \{0,1,\ldots,n\}$ and an edge set $\mathcal{E}$. Node $0$ is known as the substation, and $\mathcal{N} = \mathcal{N}_0/\{0\}$ denotes the set of nodes excluding the substation node. 
Each node $i\in\mathcal{N}$ is associated with an active power injection $p_i$ and a reactive power injection $q_i$. Let $v_i$ be the squared voltage magnitude, and let $p, q$ and $v$ denote $\{p_i,q_i,v_i\}_{i \in \mathcal{N}}$ stacked into a vector. 
The variables satisfy the following equations, $\forall i \in\mathcal{N}$, 

\vspace{-0.4cm}
 \small
\begin{subequations}\label{eq:linear_distflow}
{
\begin{align}
    p_i &= -P_{ji}  + \sum_{k: (i, k) \in \mathcal{E}} P_{ik}\,, \quad
    q_i = -Q_{ji} + \sum_{k: (i, k) \in \mathcal{E}} Q_{ik}\,, \label{eq:conserv_law}\\
    v_i &= v_j - 2(r_{ji}P_{ji} + x_{ji} Q_{ji})\,, \label{eq:bfm_vl}
\end{align}}
\end{subequations}
\normalsize
where $j$ is the parent node of $i$ in the distribution network, $P_{ji}$ and $Q_{ji}$ represent the active power and reactive power flow on line $(j,i)$, and $r_{ji}$ and $x_{ji}$ are the line resistance and reactance. 
\eqref{eq:linear_distflow} can be written in the vector form,
\small
\begin{equation}\label{eq:power_flow}
v = R p + X q + v_0 \mathbf{1} = Xq + v^{env} ,
\end{equation}
\normalsize
where $v^{env} = R p +v_0 \mathbf{1}$ is the non-controllable part. $R={[R_{ij}]}_{n \times n}, X = {[X_{ij}]}_{n \times n}$ are defined as
$R_{ij}:= 2 \sum_{(h, k) \in \mathcal{P}_i \cap \mathcal{P}_j} r_{hk}$, $X_{ij}:= 2 \sum_{(h, k) \in \mathcal{P}_i \cap \mathcal{P}_j} x_{hk}$. Here, $\mathcal{P}_i$ is the set of lines on the unique path from bus $0$ to bus $i$, and $v_0$ is the squared
voltage magnitude at the substation bus. $R$ and $X$ are positive definite matrices and all elements are positive \cite{shi2022stability}.

We make the following assumptions that are well-justified for voltage control on distribution networks~\cite{7028508, shi2022stability}.
\begin{assumption}
The system models (i.e., matrices $R$ and $X$ in \eqref{eq:power_flow}) are time-invariant, and the controllers are installed in every bus without real-time communication.  
\end{assumption}
\begin{assumption}
   There is a timescale separation between the voltage dynamics and the dynamics of inverters, so that the controlled inverter  injects instantaneously the exact value of reactive power computed by the controller. 
\end{assumption}

\subsection{Voltage Control Problem}
The optimal voltage control problem at the steady state is,
\vspace{-0.1cm}
\small
\begin{subequations}
\label{opt:steady1}
\begin{align}
\min_{q} \quad & F(q)=C(q)+\frac{1}{2}q^\top Xq+q^\top \Delta\Tilde{v} \label{eq_ss:obj}\\
\text{s.t.}\quad & \underline{q} \leq q \leq \bar{q}
\end{align}
\end{subequations}
\normalsize
where $C(q)$ is the control cost and $\Delta\Tilde{v}=v^{env}-v^{nom}$. We define the reactive power safety set as $\mathcal{S}_q=\{q\in \mathbb{R}^n|\underline{q} \leq q \leq \bar{q}\}$. 
For the per-unit system, we define $v^{nom}=1$ p.u. 
Using \eqref{eq:power_flow}, the objective function can be rewritten as $F(q) = C(q) + \frac{1}{2}(v-v^{nom})^\top X^{-1}(v-v^{nom}) - \frac{1}{2} \Delta \tilde{v}^\top X^{-1} \Delta \tilde{v}$. Since the last term is a constant, the objective function finds an optimal trade-off between minimizing the control cost $C(q)$ and the voltage deviation $\frac{1}{2}(v-v^{nom})^\top X^{-1}(v-v^{nom})$. Following \cite{qu2019optimal}, we consider $C(q)=\sum_i^n C_i(q_i)$, $C_i(q_i)=\frac{\eta_i}{2{\bar{s}_i}} q_i^2$, where $\eta_i, {\bar{s}_i} > 0$ represent the cost of reactive power of bus $i$ and its apparent power capacity. Compactly, $C(q)=\frac{1}{2}q^\top C_q q$, where $C_q$ is the diagonal matrix $diag\{\frac{\eta_i}{{\bar{s}_i}}\}_{i\in\mathcal{N}}$. 

Note that the objective function~\eqref{eq_ss:obj} can be equivalently written as the sum of cost at all nodes by \eqref{eq:power_flow},

\vspace{-0.4cm}
\small
\begin{align*}
    F(q)
    =\sum_i^n \Big( C_i(q_i)+\frac{1}{2} \big( q_i(v_i+v_i^{env}-2v_i^{nom}) \big) \label{eq:decentralized_F} \Big) .
\end{align*}
\normalsize

The gradient of the objective function $\nabla F$ is,
\vspace{-0.1cm}
\small
\begin{equation}
\label{eq:gradient}
 \nabla F=C_q q + Xq+\Delta\Tilde{v} \stackrel{\circled{1}}{=} C_q q+ v - v^{nom}.
\end{equation}
\normalsize
where $\circled{1}$ follows from \eqref{eq:power_flow} and the definition of $\Delta\Tilde{v}:=v^{env}-v^{nom}$. We write $\nabla F_i=\frac{\eta_i}{{\bar{s}_i}}q_i+v_i-v_i^{nom}.$ The decomposable structure of the objective function and the gradient enables decentralized training and deployment of a controller.
We make the assumption that the optimal solution of \eqref{opt:steady1} is unique and the corresponding voltage $v^*$ lies in the safe voltage range.

\begin{assumption}\label{asp2} 
    The steady-state optimization problem \eqref{opt:steady1} is strictly convex, 
    the optimal solution $(v^{*}, q^{*})$ satisfies $v^{*} \in \mathcal{S}_v=\{ v\in\mathbb{R}^n: \underline{v}_i\leq v_i \leq\bar{v}_i \}$, $q^{*} \in \mathcal{S}_q$ where $\underline{v}_i, \bar{v}_i$ are upper and lower bounds of desired system voltage magnitudes.  
\end{assumption}
 
To solve \eqref{opt:steady1}, \cite{7436388} introduces a projected gradient method 
\small
\begin{equation}
    \label{ctl:steven'follow}
    q_i(t+1)=[q_i(t)-\gamma \nabla F_i]_{\underline{q}_i}^{\bar{q}_i} ,
\end{equation}
\normalsize
where $[\cdot]^a_b$ denotes the projection onto $[a,b]$, and $\underline{q}_i$ and $\bar{q}_i$ are the lower and upper bound of reactive power capacity. If the stepsize $\gamma$ satisfies $\gamma<\frac{2}{\lambda_{max}(\nabla^2C(q)+X)}$, where $\lambda_{max}$ denotes the largest eigenvalue, $v(t)$ and $q(t)$ under the controller (\ref{ctl:steven'follow}) converge to $(v^*, q^*)$ -- the optimal solution of (\ref{opt:steady1}).  However, this approach does not account for optimizing transient performance, which is critical when the system is subject to rapid voltage deviations due to renewable integration and EV charging (reflected in changes in $v^{env}$). This limitation motivates our design of a controller that jointly optimizes both steady-state and transient performance.

\section{Joint Transient and Steady State Performance Optimization}
In this section, we first introduce the joint transient and steady-state optimization problem. Then, we propose a \textit{Transient and Steady-state Reinforcement Learning} (TASRL) framework to solve it. 
We provide transient stability and steady-state optimality guarantees in the next section. 

From the system operator perspective, we wish to achieve two main objectives: 1) \emph{transient stability and performance:} fast convergence of system voltage to the desired operating range $\mathcal{S}_v$ (e.g., $\pm 5$\% around $v^{nom}$) after a disturbance; 2) \emph{steady-state performance:} maintaining the system operation at the most economical point. Thus, the joint transient and steady-state optimization problem is formulated as,

\vspace{-0.4cm}
\small
\begin{subequations}
\label{opt:trans_ctrl}
\begin{align}
\min_{\mathbf{\theta}} \quad & J(\theta)= \int_{t=0}^{t_f} \gamma^t \sum_{i=1}^n c_i({v}_i(t), q_i(t))  \label{eq_rl:obj}\\
\text{s.t.}\quad 
& v(t) = Xq(t) + v^{env} \,,\\
    &\dot{q}(t)=f_{\theta}(q(t), v(t))\,,\\
& v^* = \lim_{t \rightarrow t_f} v(t), q^* =\lim_{t \rightarrow t_f} q(t)\,, \label{eq:converge1}\\
& q^* \text{is the optimal solution for (3)}\,, \label{eq:converge2}\\   
&q(t)\in \mathcal{S}_q\,, \forall t \text{\ and\ } v^* \in \mathcal{S}_v. \label{eq:reactive_safety}
\end{align}
\end{subequations}
\normalsize
\vspace{-0.1cm}
where $\gamma$ is the discount factor and $c_i$ is the cost function at node $i$, for which we choose
$c_i(v_i,q_i)=C_i(q_i)+\frac{1}{2}q_i(v_i+v_i^{env}-2v_i^{nom})$. $f_{\theta}(\cdot,\cdot)$ is the controller to be optimized. 
Here, 
$t_f \in \mathbb{R}_{>0}$ is the (possibly unbounded) stabilization time.
During the transient period $[0, t_f]$, the goal is to recover the voltage quickly under limited reactive power resources while minimizing the control effort. In addition, we want the system to converge to the steady-state optima $\lim_{t \rightarrow t_f} v(t) \rightarrow v^*, \lim_{t \rightarrow t_f} q(t) \rightarrow q^*$, and such that the equilibrium point $(v^*, q^*)$ solves the steady-state optimization~\eqref{opt:steady1}. 

\paragraph*{Controller synthesis inspired by safe gradient flow}
Here, we introduce the proposed decentralized controller design. Each controller measures the local voltage magnitude and computes the local reactive power injection without real-time communication.  
To jointly optimize the transient and steady-state performance, let's start by considering a direct combination of a transient policy $\pi_{\theta}(v)$, parameterized by a neural network, and the gradient of the long-term operation cost $\nabla F(q)$, cf.~\eqref{eq:gradient},

\vspace{-1mm}
\small
\begin{equation}\label{eq:unconstraied_flow}
    \dot{q} = f_{\theta}(q, v) := -\nabla F(q) + \pi_{\theta}(v) , 
\end{equation}
\normalsize
with $q(0)=0$.
The main issue with the controller~\eqref{eq:unconstraied_flow} is that the resulting trajectory of $q(t)$ may not satisfy the reactive power constraint $\underline{q} \leq q(t) \leq \overline{q}$ at all times. To enforce it, we build on the safe gradient flow framework introduced in~\cite{allibhoy2022control}. This design employs a control barrier function $g(q)$ to ensure that a given  dynamics never leaves a safe set $\mathcal{S}_q=\{q\in \mathbb{R}^n|g(q)\leq 0\}$, where $g(q)$ is defined by  

\vspace{-4mm}
\small
\begin{align}\label{eq:g}
g(q)=\begin{bmatrix}
    I&-I
\end{bmatrix}^\top q+\begin{bmatrix}
    -\bar{q}&\underline{q}
\end{bmatrix}^\top .
\end{align}
\normalsize
If the dynamics $\dot{q} = f_{\theta}(q, v)$ satisfies
\small
\begin{equation}\label{eq:cbf_constr2}
    {\frac{\partial g}{\partial q}} f_{\theta}(q, v) \leq -\alpha g(q) ,
\end{equation}
\normalsize
then, by Nagumo’s theorem \cite{FB-SM:07},
$q(t)$ must stay inside the safe region $\mathcal{S}_q$ for all $t$. Here, the hyperparameter $\alpha>0$ indicates the degree of conservatism regarding the reactive power constraints, where the larger $\alpha$ is, the more flexibility is allowed when $q(t)$ is not reaching the constraints. 
To ensure that the controller in \eqref{eq:unconstraied_flow} satisfies the safety constraints, the safe gradient flow framework~\cite{allibhoy2022control} prescribes modifying it minimally according to the following control
barrier function quadratic program (CBF-QP) safety filter~\cite{gurriet2018towards},

\vspace{-0.4cm}
\small 
\begin{subequations}\label{eq:controller_equiv}
\begin{align}
    f_{\theta}(q, v) = \argmin_{\xi \in \mathbb{R}^n} &\frac{1}{2} \|\xi - \left(-\nabla F(q) + \pi_{\theta}(v)\right)\|_2^2 \label{eq:cbf_obj}\\
    s.t. \quad & {\frac{\partial g}{\partial q}} \xi \leq -\alpha g(q) \label{eq:cbf_constraint}
\end{align}
\end{subequations}
\normalsize 
Plugging in $g(q)$ gives,
\vspace{-4mm}

\small 
\begin{subequations}
\label{ctl:controller}
\begin{align}
f_{\theta}(q, v)=&\argmin_{\xi\in\mathbb{R}^n} \frac{1}{2}||\xi+\nabla F(q)-\pi_\theta(v)||^2\\
\text{s.t.}\quad 
    & \alpha (\underline{q}-q) \leq \xi \leq \alpha (\bar{q}-q)
    \label{ctl:condition}
\end{align}
\end{subequations}
\normalsize

The controller \eqref{ctl:controller} is our proposed controller to solve the joint optimization problem in \eqref{opt:trans_ctrl}.
We envision that (\ref{ctl:controller}) finds the safe control action closest to $-\nabla F(q)+\pi_\theta(v)$ while ensuring the reactive power constraints are never violated. If sufficient reactive power capacity exists, $f_{\theta}(q, v) = -\nabla F(q)+\pi_\theta(v)$, otherwise, the action is projected to ensure reactive power capacity constraints are met. As $\alpha\to\infty$, \eqref{ctl:controller} reduces to a projection of $-\nabla F(q(t))+\pi_\theta(v(t))$ onto $[\underline{q}, \overline{q}]$.

\begin{proposition}\label{proposition1}
The optimal solution to \eqref{ctl:controller} is given by, 
\small
\begin{equation}
\label{ctl:solution}
    f_\theta(q, v)=[\pi_{\theta}(v)-\nabla F(q)]_{\alpha(\underline{q}-q)}^{\alpha(\bar{q}-q)}
\end{equation}
\normalsize 
where $[\cdot]^a_b$ denotes the projection onto $[a,b]$. 
\end{proposition}

Proposition~\ref{proposition1} is a direct result of solving the QP~\cite[Chapter 8]{boyd2004convex} formulated by Equation \eqref{ctl:controller}.
We summarize the proposed controller in Algorithm \ref{alg:satrl}. As observed in Algorithm \ref{alg:satrl}, the controller computation and execution are decentralized. The training process of TASRL follows the same flow as standard policy optimization RL algorithms.
Each local transient policy $\pi_{i, \theta_i}(v_i)$ can be parameterized as neural networks (with requirements specified in Section \ref{sec:theory} Def. \ref{def:stable_ctl}) and trained together with $\nabla F$ 
to optimize the transient performance. The proposed TASRL framework is general and can be integrated with most policy optimization methods, including DDPG~\cite{lillicrap2015continuous}, PPO~\cite{schulman2017proximal}, and TRPO~\cite{schulman2015trust}.

\setlength{\textfloatsep}{4pt}
\begin{algorithm}[t]
	\caption{Transient and Steady-state Reinforcement Learning (TASRL) for Distribution Grid Voltage Control}
	\label{alg:satrl}
	\begin{algorithmic}
	\Ensure policy networks $\pi_{i,\theta_i}(v_i)$ with parameters $\theta_i$; hyperparameter $\alpha$; sampling time $h$; replay buffers $\mathcal{D}_i, \forall i \in \mathcal{N}$. 
	\For {$j = 0$ to $N_{ep}$}
	  \State{Randomly generate initial states $v(0)$} 
	  \For {$t = 0$ to $N_{step}$}
            \State{For each agent $i \in\mathcal{N}
            $}
	    \State{Measure the current state $v_i(t)$}
            \State{Compute the control action (reactive power adjustment)  $f_{i,\theta_i}(q_i(t), v_i(t))=[\pi_{i,\theta_i}(v_i(t))-\nabla F_i]_{\alpha(\underline{q}_i-q_i(t))}^{\alpha(\bar{q}_i-q_i(t))}$}
	    \State {Execute $q_i(t+1) = q_i(t) + hf_{i,\theta_i}(q_i(t), v_i(t))$}
            \State {Transit to next state ${v_i}(t+1)$, receive cost ${c_i}(t)$}
	    \State{Store $\{v_i(t),q_i(t), f_{i,\theta_i}(q_i(t), v_i(t)), -c_i(t),v_i(t+1)\}$ in $\mathcal{D}_i$ }
     \State{Update policy network $\theta_i$}
	  \EndFor
	\EndFor
	\end{algorithmic} 
\end{algorithm}


\section{Transient Stability and Steady-State Optimality Guarantees} 
\label{sec:theory}
In this section, we establish the closed-loop stability and optimal steady-state performance properties of Algorithm \ref{alg:satrl}.
 The guarantees rely on certain structural constraints of the transient policy in the next definition.
 
\begin{definition}[Stable decentralized transient policy]    \label{def:stable_ctl}
A set of local policy $\{\pi_{i, \theta_i}, \forall i \in \mathcal{N}\}$ is a stable transient policy if it satisfies the following conditions for each bus $i \in \mathcal{N}$:
\begin{enumerate}
\item $\pi_{i,\theta_i}(v_i)$ is a continuously differentiable function satisfying $\pi_{i,\theta_i}(v_i) = 0$ for $v_i \in [\underline{v}_i,\bar{v}_i]$;
\item $\pi_{i,\theta_i}(v_i)$ is monotonically decreasing for $v_i \in (-\infty,\underline{v}_i) \cup (\bar{v}_i,\infty)$;
\item $\pi_{i,\theta_i}(v_i)$ is bounded i.e. $c\alpha (\underline{q}^{'}_i-q_i(t))\leq \pi_{i,\theta_i}(v_i(t)) \leq c\alpha (\bar{q}^{'}_i-q_i(t))$, where $\underline{q}^{'}_i=\underline{q}_i(1-\epsilon)$, $\bar{q}^{'}_i=\bar{q}_i(1-\epsilon)$, and $\epsilon \in (0,1)$ $c\in [0,1)$.
\end{enumerate}
\end{definition}

 We write $\pi_{i,\theta_i}(v_i)=\pi_{i,\theta_i}(v_i)-\pi_{i,\theta_i}(v_i^*)$ as $\pi_{i,\theta_i}(v_i^*)=0$.  For $v_i\neq v_i^*$, we define $K_{ii}(v_i)=\frac{\pi_{i,\theta_i}(v_i)-\pi_{i,\theta_i}(v_i^*)}{v_i-v_i^*}$
 and $K(v):=-\text{diag}\left(K_{11}(v_1), K_{22}(v_2),\cdots,K_{nn}(v_n)\right)$.
By the monotonically decreasing condition in Definition~\ref{def:stable_ctl}, when $v_i \in (-\infty,\underline{v}_i) \cup (\bar{v}_i,\infty)$, $K_{ii}(v_i)<0$. When $v_i \in [\underline{v}_i,\bar{v}_i] \text{ and } v_i\neq v_i^*$, $K_{ii}(v_i)=0$. We define $K_{ii}(v_i) =0$ if $v_i=v_i^*$. As a result, for every $v$, we can write $\pi_\theta(v)=-K(v) (v-v^*)$. 
Define $\sigma_{\max}(\cdot)$ and $\sigma_{\min}(\cdot)$ as the largest and smallest singular value of a matrix. The following result establishes the theoretical guarantees.
\begin{theorem}[Transient Stability and Steady-State Optimality]\label{thm:optimality(stability)}
Suppose Assumption \ref{asp2} holds, $\pi_\theta$ is a stable decentralized transient policy according to Definition \ref{def:stable_ctl}, and $2\sigma_{\max} (K(v)) \leq \sigma_{\min}(C_qX^{-1}+I)\,, \forall v \in \mathbb{R}^n$,
then with a sufficiently large $\alpha$,
the closed-loop system is asymptotically stable with controller \eqref{ctl:controller}. In addition, $\lim\limits_{t \rightarrow t_f} v(t) = v^*, \lim\limits_{t \rightarrow t_f} q(t) = q^*$, $q^*$ is the global minimizer of optimization problem \eqref{opt:steady1}, $v^*\in \mathcal{S}_v$, and $q(t)\in \mathcal{S}_q\,, \forall t\geq 0$. 
\end{theorem}
Theorem \ref{thm:optimality(stability)} shows that with a Lipschitz-like bound on the transient policy $\pi_{i, \theta_i}, \forall i \in \mathcal{N}$, the proposed controller in \eqref{ctl:controller} obtains both transient stability and steady-state optimality while respecting the reactive power capacity constraint at all times. Below, we present the theoretical analysis of Theorem~\ref{thm:optimality(stability)}.
\begin{lemma}\label{lemma:1}
Suppose $2\sigma_{\max} (K(v)) \leq \sigma_{\min}(C_qX^{-1}+I), \forall v \in \mathbb{R}^n$ and let $\langle \cdot, \cdot\rangle$ be the dot product, then $\lVert -\nabla F(q)\rVert^2+2\langle \pi_\theta(v),-\nabla F(q)\rangle\geq 0\,, \forall v \in \mathbb{R}^n, q \in \mathcal{S}_q$.
\end{lemma}
\vspace{-0.4cm}

\begin{proof}
Following \eqref{eq:gradient} and \eqref{eq:power_flow}, $\nabla F(q)= \nabla F(q)-\nabla F(q^*)=(C_q X^{-1} + I)(v-v^*)$. Denote $A=C_q X^{-1} + I$, we have 

\vspace{-4mm}
\small
\begin{align*}
    &\quad \lVert -\nabla F(q)\rVert^2+2\langle \pi_\theta(v),-\nabla F(q)\rangle\\
    &=(v-v^*)^\top\left [A^\top A+K(v)A+A^\top K(v)\right](v-v^*).
\end{align*}
\normalsize
To ensure $\lVert -\nabla F(q)\rVert^2+2\langle \pi_\theta(v),-\nabla F(q)\rangle\geq 0$, it suffices to have $A^\top A+K(v)A+A^\top K(v) \succeq 0$.
\vspace{-4mm}

\small
\begin{subequations}
    \begin{align}
    &\quad A^\top A+K(v)A+A^\top K(v) \succeq 0\\
    &\iff [A+K(v)]^\top[A+K(v)] \succeq K(v)^\top K(v)\\
    &\iff \sigma_{\min}(A+K(v))\geq \sigma_{\max}(K(v)) \label{ineq:singular}
\end{align}
\end{subequations}
\normalsize
\vspace{-0.4cm}

By \cite[Proposition 9.6.8]{10.2307/j.ctt7t833.15}, $\sigma_{\min}(A+K(v))\geq \sigma_{\min}(A)-\sigma_{\max}(K(v))$. Thus, $\sigma_{\min}(A)\geq 2\sigma_{\max} (K(v))$ is a sufficient condition for $\lVert -\nabla F(q)\rVert^2+2\langle \pi_\theta(v),-\nabla F(q)\rangle\geq 0$.
\end{proof}
\vspace{-0.4cm}
\subsection{Proof of Theorem~\ref{thm:optimality(stability)}}
\label{sec:proof}
\begin{proof}[Proof of Theorem~\ref{thm:optimality(stability)}]
By design, the proposed controller~\eqref{ctl:controller} guarantees $q(t)\in \mathcal{S}_q$, $\forall t\geq 0$ by the CBF-QP safety filter. We will work with the equivalent controller form in \eqref{eq:controller_equiv} throughout the proof since the inequality constraints are in a more compact form. The Lagrangian of \eqref{eq:controller_equiv} is,

\vspace{-0.4cm}
\small
\begin{align*}
L(\mathbf{\xi},\mathbf{\omega};q)=&\frac{1}{2}\lVert\mathbf{\xi}+\nabla F(q)-\pi_\theta(v)\rVert_2^2+\mathbf{\omega}^\top \Big(\frac{\partial g(q)}{\partial q}\xi +\alpha g(q) \Big) ,
\end{align*}
\normalsize 
where $\omega$ is the nonnegative Lagrange multiplier for \eqref{eq:cbf_constraint}.
The \textit{Karash-Kuhn-Tucker} (KKT) conditions~\cite[Chapter~3]{Bertsekas/99} of (\ref{eq:controller_equiv}) are:

\vspace{-4mm}
\small
\begin{subequations}
\label{opt:KKT}
\begin{align}
\mathbf{\xi}+\nabla F(q) -\pi_\theta(v)+ \frac{\partial g(q)}{\partial q}^\top\mathbf{\omega}&=0 \label{eq:lagrang}\\
\mathbf{\omega}\geq 0\,, \frac{\partial g(q)}{\partial q}\xi +\alpha g(q)\leq 0\\
\mathbf{\omega}^\top \left(\frac{\partial g(q)}{\partial q}\xi +\alpha g(q)\right)=0\label{eq:lagrang-d}
\end{align}
\end{subequations}
\normalsize
Because \eqref{eq:controller_equiv} is strongly 
convex with respect to $\xi$, the existence of a $(\mathbf{\xi},\mathbf{\omega})$ satisfying (\ref{opt:KKT}) is sufficient for $\mathbf{\xi}=f_\theta(q)$. To verify the feasibility of the KKT conditions, we apply the Mangasarian-Fromovitz Constraint Qualification (MFCQ) condition~\cite[Chapter~3]{Bertsekas/99}, which requires a $\mathbf{\xi}\in\mathbb{R}^n$ s.t.

\vspace{-0.2cm}
\small
\begin{equation*}
    \nabla g_i(q)^T\xi<0 \quad \forall i\in I_0(q)=\{1\leq i\leq 2n|g_i(q)=0\}
\end{equation*}
\normalsize
where $I_0(q)$ is the active constraint set. Given the specific structure of $g(q)$, $\forall q\in\mathcal{S}_q$, there always exists an $\xi$ such that the MFCQ is satisfied. By Lemma 4.5 of \cite{allibhoy2022control}, the existence of a solution $(\mathbf{\xi},\mathbf{\omega})$ satisfying \eqref{opt:KKT} is guaranteed. We next characterize the stability properties of our proposed algorithm with the solution $(\mathbf{\xi},\mathbf{\omega})$ of the KKT conditions \eqref{opt:KKT}.

An immediate result of Lemma \ref{lemma:1} is 

\vspace{-0.5cm}
\small
\begin{align*}
    &\lVert \pi_\theta(v)-\nabla F(q)\rVert^2=\langle\pi_\theta(v)-\nabla F(q),\pi_\theta(v)-\nabla F(q) \rangle\\
    &=\lVert \pi_\theta(v)\rVert^2+\lVert -\nabla F(q)\rVert^2+2\langle \pi_\theta(v),-\nabla F(q)\rangle \geq \lVert \pi_\theta(v)\rVert^2
\end{align*}
\normalsize 
For every bus $i$ such that $\pi_{i,\theta_i}(v_i)\neq 0$, we have $|[\pi_{i,\theta_i}(v_i)-\nabla F_i(q)]_{\alpha(\underline{q}_i-q_i)}^{\alpha(\bar{q}_i-q_i)}|\geq |[\pi_{i,\theta_i}(v_i)-\nabla F_i(q)]_{\alpha\epsilon(\underline{q}_i)}^{\alpha\epsilon(\bar{q}_i)}|$. Given that $-\nabla F_i(q)$ is bounded on both sides, there exists a finite $\alpha$ such that $[\pi_{i,\theta_i}(v_i)-\nabla F_i(q)]_{\alpha(\underline{q}-q)}^{\alpha(\bar{q}-q)}=\pi_{\theta}(v_i)-\nabla F_i(q)$. A similar reasoning holds if $\pi_{i,\theta_i}(v_i) = 0$. Therefore, with a sufficiently large $\alpha$, we have  $\lVert f_\theta(q, v) \rVert = \lVert [\pi_{\theta}(v)-\nabla F(q)]_{\alpha(\underline{q}-q)}^{\alpha(\bar{q}-q)} \rVert  \geq \lVert \pi_{\theta}(v) \rVert.$  
Using $F(q)$ as a Lyapunov-like function, from Eq \eqref{eq:lagrang}, we have  $\nabla F(q) =-\frac{\partial g(q)}{\partial q}^\top\mathbf{\omega}  -f_\theta(q, v)+\pi_\theta(v)$, thus
\small
\begin{align}
\label{opt:growth}
&L_{f_\theta}F(q(t))=f_\theta(q(t), v(t))^\top\nabla F(q(t)) \nonumber\\
&=f_\theta(q(t), v(t))^\top \left(-\frac{\partial g(q(t))^\top}{\partial q(t)}\mathbf{\omega}-f_\theta(q(t), v(t))+\pi_\theta(v(t))\right) \nonumber\\
&=-\lVert f_\theta(q(t), v(t))\rVert^2+f_\theta^\top(q(t), v(t))\pi_\theta(v(t))+\alpha \mathbf{\omega}^\top g(q(t)) \nonumber\\
&\leq -\lVert f_\theta(q(t), v(t))\rVert^2+||f_\theta(q(t), v(t))\rVert\lVert \pi_\theta(v(t))\rVert+\alpha \mathbf{\omega}^\top g(q(t))\nonumber\\
&\leq 0
\end{align}
\normalsize
where the second equality follows Eq \eqref{eq:lagrang-d}. Given that $g(q(t))\leq 0$, $\omega$ is a nonnegative dual variable, $\alpha>0$, $\alpha\mathbf{\omega}^\top g(q(t))\leq 0$ holds, which leads to the final inequality. 

Furthermore, $L_{f_\theta}F(q)=0$ if and only if $f_\theta(q^*,v^*)=0$. Then there exists $(0,\omega^*)$, which is the solution of \eqref{opt:KKT}. Plug $(0,\omega^*)$ into \eqref{opt:KKT}, it is reduced to

\vspace{-0.5cm}
\small
\begin{subequations}
\label{opt:KKT_eq}
\begin{align}
\nabla F(q^*) -\pi_\theta(v^*)+ \frac{\partial g(q^*)}{\partial q}^\top\omega^*&=0 \label{eq:lagrang2}\\
\omega^*\geq 0\,, \alpha g(q^*)\leq 0\\
(\omega^*)^\top \left(\alpha g(q^*)\right )=0\label{eq:lagrang-d2}
\end{align}
\end{subequations}
\normalsize
By Assumption \ref{asp2} where $v^*\in \mathcal{S}_v$, we have $\pi_\theta(v^*)=0$. Given that $\alpha>0$, it follows immediately that $(v^*,q^*)$ is the optimal solution of \eqref{opt:steady1}. 
Due to the strict convexity of $F(q)$, $q^*$ is the unique global minimizer. By Lyapunov Stability Theory \cite[Chapter~3]{Khalil:1173048}, we conclude that the closed loop system is asymptotically stable with respect to the global minimizer. 
\end{proof}
\vspace{-0.7cm}
\subsection{Stable transient policy design}
We now present the neural network design that meets the stable decentralized transient policy $\pi_{i,\theta_i}(v_i)$ in Definition~\ref{def:stable_ctl}.

\textbf{Conditions 1) and 2):} 
To ensure conditions 1) and 2), we adopt the structure in~\cite{cui2022structured} for each bus $i$. The stacked ReLU function constructed by Eq~\eqref{eq:relu_pos} is monotonically decreasing for $v_i-\bar{v} > 0$ and zero when $v_i-\bar{v} \leq 0$. 

\vspace{-0.3cm}
\begin{small}
\begin{subequations}\label{eq:relu_pos}
    \begin{align}
    &\xi^{+}(v_i-\bar{v}; w^+, b^+) = {(w^+)^\top} \text{ReLU}(\mathbf{1} (v_i-\bar{v}) + b^+),\\
    &\sum_{l=1}^{d'} w^{+}_l < 0, \forall d' = 1, \!\cdots\!, d\,, b^{+}_1 = 0, b^{+}_l \leq b^{+}_{l\!-\!1}, \forall l =2, \cdots, d. 
    \end{align}
\end{subequations}
\end{small}
The stacked ReLU function constructed by Eq~\eqref{eq:relu_neg} is monotonically decreasing for $v_i-\underline{v} < 0$ and zero otherwise.

\vspace{-0.4cm}
\begin{small}
\begin{subequations}\label{eq:relu_neg}
    \begin{align}
    &\xi^{-}(v_i-\underline{v}; w^{-}, b^{-}) = (w^{-})^\top \text{ReLU}(-\mathbf{1} (v_i-\underline{v}) + b^{-}),\\
    & \sum_{l=1}^{d'} w^{-}_l > 0, \forall d' = 1, \!\cdots\!, d\,, b^{-}_1 = 0, b^{-}_l \leq b^{-}_{l-1}, \forall l =2,\!\cdots\!, d. 
    \end{align}
\end{subequations}
\end{small}

\textbf{Condition 3)} This condition requires the output of $\pi_{i,\theta_i}(v_i)$ to be bounded. We use $\tanh$ activation function to scale the output as a percentage
while preserving its sign. Then the percentage is multiplied by the absolute value of the bounds. The local policy network of bus $i$ is defined as,
\vspace{-0.2cm}
\small
\begin{multline}
     \pi_{i,\theta_i}(v_i)=c\alpha(q_i-\underline{q}^{'}_i)\tanh(\xi^{+}(v_i-\bar{v}; w^+, b^+))\\
     +c\alpha(\bar{q}_i^{'}-q_i)\tanh(\xi^{-}(v_i-\underline{v}; w^{-}, b^{-})).
\end{multline}
\normalsize
It is noteworthy that the monotonicity of the $\tanh$ function ensures that conditions 1) and 2) are still satisfied. 

\section{Experiments}
In this section, we demonstrate the effectiveness of the proposed method in two IEEE distribution test systems.
\subsection{Experiment Setup}
We evaluate our approach on the IEEE 13-bus and 123-bus test feeders~\cite{8063903}. The nominal voltage magnitude for both environments at each bus except the substation is 4.16 kV, and the acceptable range of operation is $\pm 5\%$ of the nominal value, which is $[3.952\text{kV}, 4.368\text{kV}]$. Though our theoretical analysis is based on the linearized system model in \eqref{eq:power_flow}, all experiments are run using Pandapower\cite{pandapower} as the nonlinear power flow simulator to evaluate the algorithm performance. 
We simulate two different voltage disturbance scenarios: 1) \emph{High voltages:} with abundant sunshine during daylight, the PV generators generate excessive power that can lead to high voltage issues. 2) \emph{Low voltages:} the system is serving peak loads without enough generation. 
For each scenario, we vary $v^{env}$ to obtain different degrees of initial voltage disturbance, i.e., $5\%$ to $15\%$ of $v^{nom}$. 
We test two baselines:
\begin{enumerate}
    \item Stable-DDPG~\cite{shi2022stability} with  Safety Filter: The Stable-DDPG~\cite{shi2022stability} optimizes the transient performance with a stability guarantee. To enforce reactive power safety, we incorporate the CBF-QP safety filter in \eqref{eq:controller_equiv} by replacing the obj \eqref{eq:cbf_obj} with $\argmin_{\xi \in \mathbb{R}^n} \frac{1}{2} \|\xi - \pi_{\theta}(v)\|_2^2 $.
    \item Safe gradient flow~\cite{allibhoy2022control}: Safe gradient flow optimizes the steady-state performance with reactive power safety.
\end{enumerate}
We use the DDPG framework~\cite{lillicrap2015continuous} to train the policy network update in our TASRL algorithm.
While the theoretical analysis is done in continuous time, in the numerical simulation, the control is executed in a discrete manner. With $h$ denoting the sampling period, the update law of reactive power is defined as $q_i(t+1)=q_i(t)+hf_{i, \theta_i}(q_i(t), v_i(t))$.
For all experiments, we use $h=1 s$, $\alpha=0.5$ $t_f=100 s$,  and $\gamma=0.99$.
Following \cite{5768094}, we set $\bar{q}=\sqrt{\bar{s}^2-\bar{p}^2}\approx 0.45\bar{p}$, $\underline{q}=-\bar{q}$.  
For both test cases, instead of enforcing $2\sigma_{\max} (K(v)) \leq \sigma_{\min}(C_qX^{-1}+I)$ during training, we verify the trained controller with 1,000 random control trajectories with 500 time steps / traj. For all the sampled values of $v$, we verified
the stability condition holds with matrix $X$ from the simulated model.

\subsection{Results}
\begin{figure}[tb]
    \centering
    \includegraphics[width=0.21\textwidth]{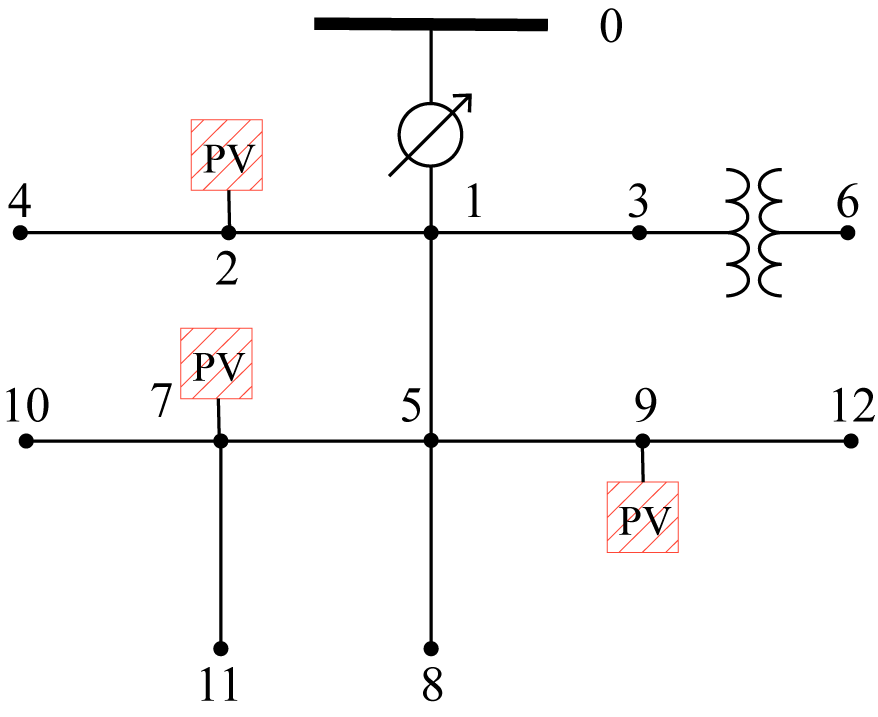}
    \includegraphics[width=0.22\textwidth]{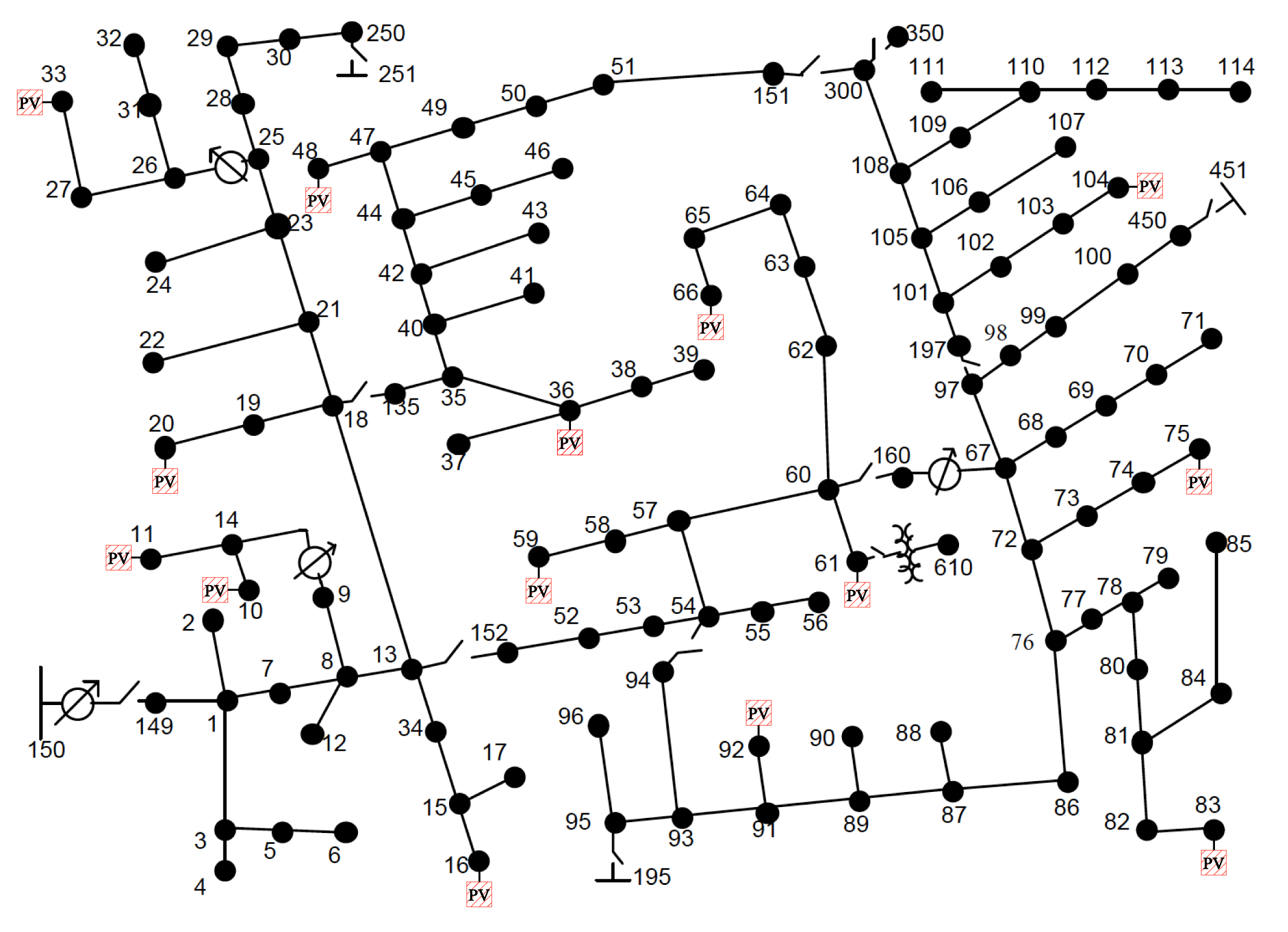}
    \caption{Left: Schematic diagram of the 13 bus system with three PV generators and voltage controllers located at nodes 2, 7, and 9. Right: Schematic diagram of 123 bus system, with 14 PV generators and voltage controllers located at nodes 10, 11, 16, 20, 33, 36, 48, 59, 61, 66, 75, 83, 92, and 104.}
    \label{fig:13_123bus}
\end{figure}
\subsubsection{IEEE 13-bus}
IEEE 13-bus system is a standard radial distribution system depicted in Figure \ref{fig:13_123bus} (Left), where three PV stations and controllers are located at buses 2, 7, and 9. 

Table \ref{table:13_bus_result} compares the transient and steady-state performance of 500 different voltage violation scenarios. Clearly, TASRL achieves the best performance for both transient and steady states. 
For this test case, the magnitude of the gradient is relatively large. 
As a result, the transient performance of the safe gradient flow is close to TASRL, and the Stable-DDPG is underperforming. 
In terms of steady-state performance, both the safe gradient flow and the TASRL achieve the best result as $v(t)$ and $q(t)$ converge to the steady-state optima~$(v^*, q^*)$. 
\begin{table}[tb]
\centering
\caption{Performance of 500 scenarios for 13-bus system.}
 \label{table:13_bus_result}
 \begin{tabular}{lccc}
    \toprule
     & \multicolumn{2}{c}{Transient Performance}  & \multicolumn{1}{c}{Steady State}  \\
    \cmidrule(r){2-4}
     Method& \shortstack{Recovery\\Time (s)}  & \shortstack{Transient\\Cost} & \shortstack{Objective\\ $F(\mathbf{q})$ } \\
    \midrule
   Stable-DDPG \cite{shi2022stability}& 10.18 
    & -5.61 
    & -0.09\\
    Safe gradient flow \cite{allibhoy2022control} &3.04  & -6.74 & 
    \textbf{-0.11}  \\
    TASRL & \textbf{2.60} & \textbf{-6.76} & \textbf{-0.11}\\
    \bottomrule
\end{tabular}
\\Note: Smaller value means better performance for all three columns.
\end{table}
\subsubsection{IEEE 123-bus}
Figure \ref{fig:13_123bus} (Right) demonstrates the IEEE 123-bus distribution test feeder, which has 14 PV generators and controllers randomly placed in the network. 
We summarize the performance of our method and the baselines in Table \ref{table:123_bus_result}. The average response time for TASRL is 12.08 steps, which saved $77\%$ of time compared to the safe gradient flow, and $30\%$ compared to the Stable-DDPG. Both the TASRL and safe gradient flow obtain optimal steady-state cost. 
Interestingly, compared to the results of IEEE 13-bus, the gap between the steady-state performance of Stable-DDPG and the other two methods is larger. This indicates that optimizing the steady-state performance becomes increasingly crucial as the system complexity increases.

\begin{table}[tb]
\centering
\caption{Performance of 500 scenarios for 123-bus system.}
 \label{table:123_bus_result}
 \begin{tabular}{lccc}
    \toprule
     & \multicolumn{2}{c}{Transient Performance}  & \multicolumn{1}{c}{Steady State}  \\
    \cmidrule(r){2-4}
    Method & \shortstack{Recovery\\Time (s)}  & \shortstack{Transient\\Cost} & \shortstack{Objective\\ $F(\mathbf{q})$}\\
    \midrule
    Stable-DDPG & 17.36  
    &  -303.44 
    & -4.90\\
    Safe gradient flow &52.43  & -254.72  & 
    \textbf{-5.95}  \\
    TASRL & \textbf{12.08}  & \textbf{-333.03} & \textbf{-5.95} \\
    \bottomrule
\end{tabular}
\\Note: Smaller value means better performance for all three columns.
\end{table}

\begin{figure}[htbp]
    \centering
    \includegraphics[width=8.5cm]{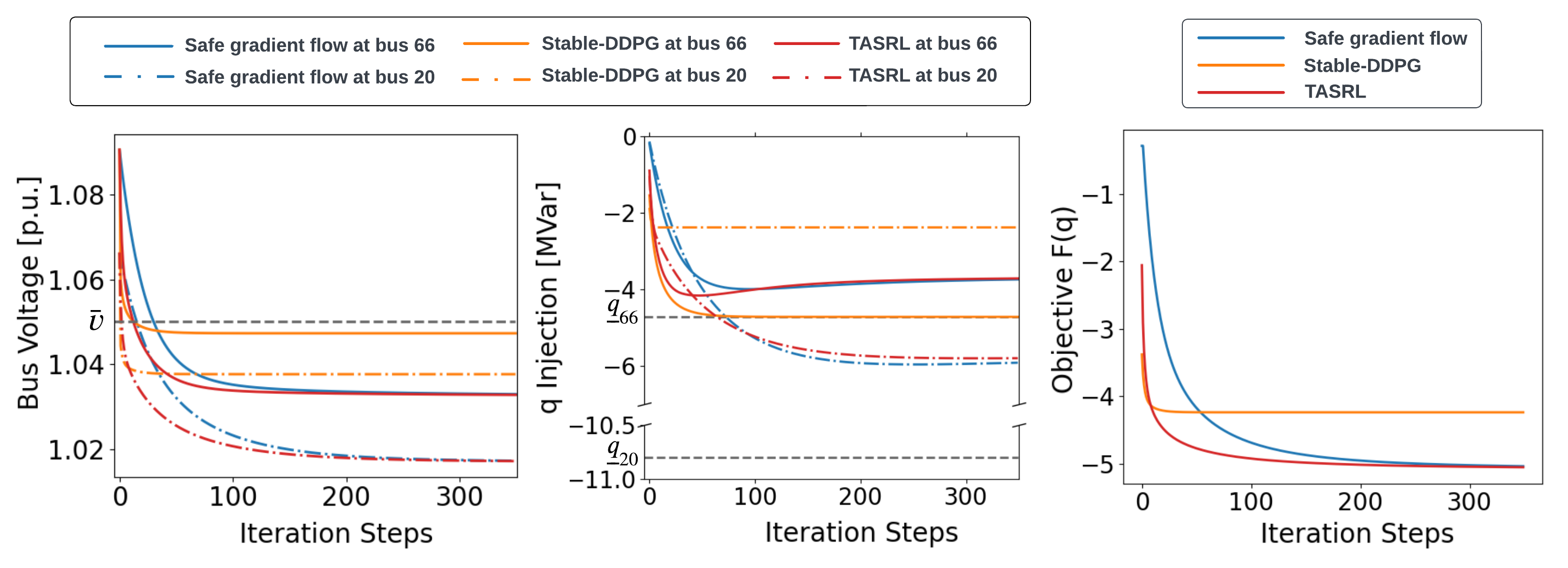}
    \caption{Control trajectory of IEEE 123-bus system. The voltage trajectory is shown in the left plot, the middle plot displays the reactive power usage, and the right plot shows the objective function's trajectory.
    }
    \label{fig:123_traj}
\end{figure}

Figure~\ref{fig:123_traj} shows an example control trajectory of the proposed approach and the baselines at bus 20 and bus 66.
Although both the Stable-DDPG and the proposed method restore voltage quickly, Stable-DDPG uses more reactive power at bus 66 and less at bus 20, leading to a suboptimal solution for $F(q)$. On the other hand, the safe gradient flow and the proposed method converge to the same steady state, while the safe gradient flow's convergence is slower.
\vspace{-1.5em}
\subsection{Effect of design parameter of safe gradient flow}
\begin{figure}[htbp]
    \centering
    \includegraphics[width=8.4cm]{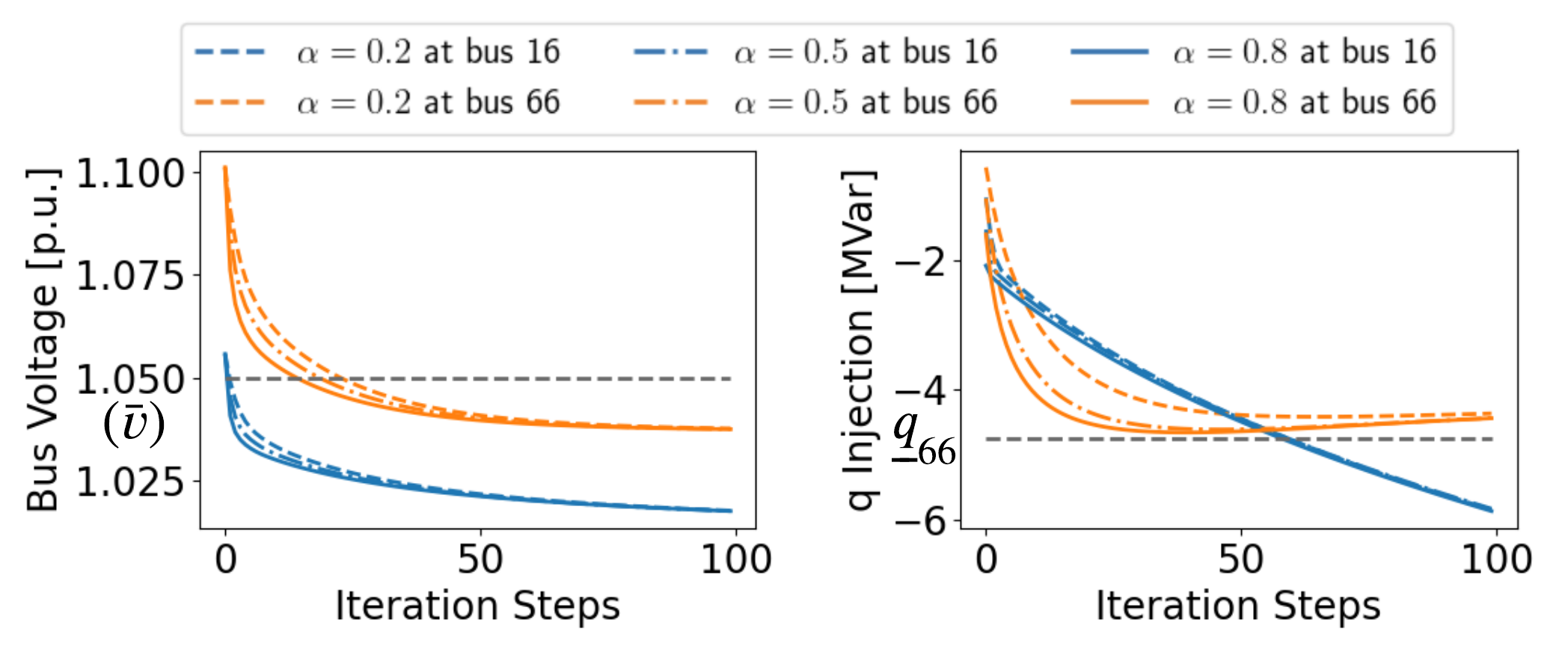}
    \caption{Control trajectory of IEEE 123-bus system with our proposed method using different $\alpha_h$. For the sake of simplicity, we did not plot $\underline{q}_{16}=-21.6$.}
    \label{fig:alpha}
\end{figure}

We illustrate the effect of hyperparameter $\alpha$ 
in Figure \ref{fig:alpha}. Smaller $\alpha$ results in a smoother voltage trajectory, as the controller becomes more conservative for the reactive power capacity constraints. The right plot shows the corresponding reactive power injection, which is not significantly affected by $\alpha$ when it is far from the capacity limit due to the presence of a transient performance optimizer (bus 16). However, when approaching the capacity limit, smaller $\alpha$ will slow down the rate of change of reactive power injection (bus 66).
\vspace{-3mm}
\vspace{-1em}

\section{Conclusions}
We proposed the TASRL framework to optimize transient and steady-state performance simultaneously for voltage control and established formal guarantees for it.
The main insight underlying our approach is that, by synthesizing a stable transient policy and a steady-state optimizer within a safe gradient flow framework, the performance of different time scales can be optimized end-to-end. Our proposed method was tested on both IEEE 13-bus and 123-bus systems. The results demonstrate that TASRL not only converges to the steady-state optimal solution but also exhibits superior transient performance compared to existing methods. Future work will (i) extend the theoretical analysis to include the controller dynamics and nonlinear system models and (ii) generalize our design to handle time-varying and dynamic loads while maintaining the stability guarantees.

\addtolength{\textheight}{-12cm}  








\end{document}